# Chiral liquid crystal colloids

Ye Yuan[1#], Angel Martinez[1#], Bohdan Senyuk[1], Mykola Tasinkevych[2,3,4] &

Ivan I. Smalyukh[1,5,6*]

[1]*Department of Physics and Soft Materials Research Center, University of Colorado, Boulder, CO 80309, USA*

[2]*Max-Planck-Institut für Intelligente Systeme, Heisenbergstr. 3, D-70569 Stuttgart, Germany*

[3]*IV. Institut für Theoretische Physik, Universität Stuttgart, Pfaffenwaldring 57, D-70569 Stuttgart, Germany*

[4]*Centro de Física Teórica e Computacional, Departamento de Física, Faculdade de Ciências, Universidade de Lisboa, Campo Grande P-1749-016 Lisboa, Portugal*

[5]*Department of Electrical, Computer, and Energy Engineering, Materials Science and Engineering Program, University of Colorado, Boulder, CO 80309, USA*

[6]*Renewable and Sustainable Energy Institute, National Renewable Energy Laboratory and University of Colorado, Boulder, CO 80309, USA*

*\*Email: ivan.smalyukh@colorado.edu*

[#]These authors contributed equally.

***Colloidal particles disturb the alignment of rod-like molecules of liquid crystals, giving rise to long-range interactions that minimize the free energy of distorted regions. Particle shape and topology are known to guide this self-assembly process. However, how chirality of colloidal inclusions affects these long-range interactions is unclear. Here we study the effects of distortions caused by chiral springs and helices on the colloidal self-organization in a nematic***

*liquid crystal using laser tweezers, particle tracking and optical imaging. We show that chirality of colloidal particles interacts with the nematic elasticity to pre-define chiral or racemic colloidal superstructures in nematic colloids. These findings are consistent with numerical modelling based on the minimization of Landau-de Gennes free energy. Our study uncovers the role of chirality in defining the mesoscopic order of liquid crystal colloids, suggesting that this feature may be a potential tool to modulate the global orientated self-organization of these systems.*

Chirality plays important roles in condensed matter and beyond[1-7], albeit understanding how the chiral symmetry breaking on the scale of individual building blocks of matter can potentially alter the physical behavior of materials[1,8] remains a challenge. For example, a well-known effect of adding chiral molecules into a nonchiral liquid crystal (LC) is that they induce chirality of the ensuing mesomorphic phase, even at vanishingly low concentrations[1], though these effects cannot be explored down to individual molecules due to the limited experimental sensitivity. The effects of adding chiral colloidal particles into a nematic LC also remain unexplored, even though they are additionally of interest from the standpoint of the colloidal paradigm[8,9] that reveals unprecedented richness of novel physical behavior and promises a host of technological applications[10-13]. When embedded within the LC, colloidal particles distort the long-range ordering of rod-like molecules[8], giving rise to distortions of the molecular alignment field that resemble the electric field configurations around various electrostatic charge distributions.[8-19] Understanding and control of self-assembly of colloidal dipoles, quadrupoles and hexadecapoles[8-18] benefit from the framework of multipole expansions,[19] which has been extended to nematic colloids[8-18]. Although geometric shape and genus of particles[20-24] allow for controlling elastic multipoles and guiding self-assembly, the role of chirality of colloidal inclusions in controlling these interactions remains unknown.

Here we develop chiral colloidal particles in the forms of left- and right-handed springs and helices suspended in a nematic LC. Using a combination of experiments and computer simulations based on the minimization of Landau-de Gennes free energy, we uncover how individual colloidal objects with well-defined chirality impart the symmetry breaking to the nematic host and how the ensuing distortions in the molecular alignment mediate chirality-dependent, elastic colloidal interactions. We explore pair interactions between like- and oppositely-handed particles with different geometric parameters, revealing unique chirality-controlled effects, which guide colloidal self-assembly. In nematic hosts with a uniform far-field alignment, the like-handed particles self-assemble into chiral colloidal structures, differently from the racemic dispersions of oppositely-handed particles. These assemblies preserve the uniform far-field alignment of the nematic host imposed by strong boundary conditions, resembling frustrated unwound chiral nematics at low concentrations of molecular chiral dopants[1,25]. Similar to how fundamental insights into the elastic interactions at the scale of individual non-chiral particles[8,9] recently led to realization of photon-upconverting triclinic pinacoidal colloidal crystals[26,27], our findings may enable enantiomorphic colloidal crystals with properties pre-engineered by controlling chiral interactions at the mesoscale.

**Individual chiral colloidal particles in a nematic host**

Using two-photon-absorption based polymerization (Supplementary Figs. 1 and 2) [6,7], we fabricate colloidal particles shaped as right- and left-handed microsprings (Fig. 1a,b). These particles set tangential boundary conditions for the director $\mathbf{n}(\mathbf{r})$, resulting in particle-induced director distortions in the LC. To minimize the energetic costs of elastic distortions, these particles align with their helical axes either (roughly) along or perpendicular to the far-field LC director $\mathbf{n}_0$ while undergoing Brownian motion (Fig. 1c,d). Colloidal diffusion is strongly anisotropic when helical axes orient roughly along $\mathbf{n}_0$ (Fig. 1c) and only weakly anisotropic when they are roughly orthogonal to $\mathbf{n}_0$ (Fig. 1d). This behavior arises from the superposition of effects due to particles' geometric shape, LC's viscosity and particle-induced defects[25-33], which exhibit orientation-dependent anisotropies that are synergistic in the former case (Fig. 1c) but compete and self-compensate in the latter case (Fig. 1d). These findings demonstrate how anisotropies of diffusion due to particle shape[28,29], LC medium's viscous properties[25,30,31] and satellite defects and director distortions[30-33], previously studied only as separate phenomena[28-33], can all interplay to enrich the colloidal behavior. Optical microscopy (Fig. 1e-k) and three-dimensional nonlinear optical imaging (Fig. 1l-n) reveal that the particle-induced director distortions mimic the chiral nature of the colloidal inclusions, as is evident from the birefringent patterns with twisting dark and bright colored brushes in the polarizing micrographs (Fig. 1f,g,i,j,k) and textures in three-photon excitation fluorescence polarizing microscopy (3PEF-PM) images (Fig. 1l-n). These experiments are consistent with the results of computational modeling based on the minimization of Landau-de Gennes free energy, $F_{LdG}$, of the LC hosting colloidal inclusions (Fig. 2). The experimental finding that the spring axes are more frequently nearly orthogonal to $\mathbf{n}_0$ (Fig. 1d,k-n) is consistent with the lower $F_{LdG}$ of this configuration (Fig. 2a). Moreover, both the modeling and experiments reveal that particles equilibrate at orientations where their helical axes tilt slightly away from $\mathbf{n}_0$ (Fig. 1c,e-j and Fig. 2a,f) or the plane orthogonal to it (Fig. 1d,k-n and Fig. 2a,h), with the tilt angles <10°. A detailed analysis of both experimental and theoretical configurations (Figs. 1 and 2) shows that this relative tilting emerges from the surface boundary conditions on the end faces of finite-length colloidal springs.

Similar to spheres, springs have Euler characteristic $\chi=2$. Following topological theorems,[22] they induce surface topological defects called "boojums", with the total strength $\sum_i s_i = \chi = 2$, where $s_i$ is the number of times the surface-projected director, $\mathbf{n}_s(\mathbf{r})$, rotates by $2\pi$ as one circumnavigates the defect core once. The theorems prescribe only the net strength of all defects but not their number. Minimization of the elastic free energy often dictates the presence of additional pairs of self-compensating defects[22]. This explains why $F_{LdG}$ versus the angle $\theta$ between the spring's axis and $\mathbf{n}_0$ (Fig. 2a) has two branches. The small-$\theta$ branch (bottom left inset of Fig. 2a) corresponds to the configurations with two $s=1$ boojums located at the edges of the spring's two end faces (inset in Fig. 2g, and Fig. 2k-m). There are four localized regions with reduced scalar order parameter altogether, though two of them are topologically trivial with $s=0$ (Fig. 2j-m and Supplementary Fig. 3). These boojums have core structures shaped as small half-integer semi-loops (handles)[22]. The large-$\theta$ branch (top-right inset of Fig. 2a) corresponds to several additional pairs of boojums at different locations on the particle surface (Fig. 2h,i,n-t). Despite of these extra self-compensating defects, this branch of the angular dependence contains the ground state at $\theta \approx 90°$. The defect pairs (Fig. 2h,i,n-t) efficiently ease the energetically "expensive" distortions of $\mathbf{n}(\mathbf{r})$ present in the configurations of the small-$\theta$ branch, thereby reducing the overall $F_{LdG}$ (Fig. 2a). The energetic cost of the nucleation of defect pairs stabilizes the $\theta \approx 0°$ metastable state against realignment into the $\theta \approx 90°$ ground state.

$F_{LdG}(\theta)$ is sensitive to the tilting direction of the helical axis, which is due to the finite length of chiral particles. For an infinitely long spring, the free energy would be degenerate with respect to tilting direction of the helical axis, but the presence of the particle ends lifts this degeneracy, further enriching behavior of our colloidal particles. To better visualize the particle orientations, we rigidly attach a right-handed orthonormal reference frame ($\hat{\mathbf{e}}_{\parallel},\hat{\mathbf{e}}_{\perp},\hat{\mathbf{e}}_3$) to each particle, as depicted in Fig. 2b-e and Supplementary Fig. 1. The axes $\hat{\mathbf{e}}_{\parallel}$ and $\hat{\mathbf{e}}_{\perp}$ are parallel and perpendicular to the helical axis of the particle, respectively, with $\hat{\mathbf{e}}_{\perp}$ chosen so that the plane spanned by $\hat{\mathbf{e}}_{\parallel}$ and $\hat{\mathbf{e}}_{\perp}$ roughly contains the spring's two end faces. We then define $\hat{\mathbf{e}}_3=\hat{\mathbf{e}}_{\parallel}\times\hat{\mathbf{e}}_{\perp}$. The dependence of $F_{LdG}$ on the tilting directions is well pronounced, especially for the large-$\theta$ branches (compare open and solid symbols in Fig. 2a).

When $\hat{\mathbf{e}}_\parallel$ rotates about $\hat{\mathbf{e}}_3$ (Fig. 2b,c), the end faces remain roughly parallel to $\mathbf{n}_0$ for all $\theta$, and in this case numerical calculations predict (almost) identical $F_{LdG}(\theta)$ for particles of right or left handedness (open symbols in Fig. 2a). However, when $\hat{\mathbf{e}}_\parallel$ tilts about $\hat{\mathbf{e}}_\perp$ (Fig. 2d,e), the end faces form an angle $\approx \theta$ with $\mathbf{n}_0$. Since the faces impose tangential anchoring on $\mathbf{n}(\mathbf{r})$, their misalignment with $\mathbf{n}_0$ is penalized by additional elastic free energy costs, yielding $F_{LdG}(\theta)$ dependence shown with solid symbols in Fig. 2a.

We also fabricate more slender left- and right-handed colloidal objects that we call "helices" (Fig. 3). Their behavior in the LC is qualitatively similar to that of the colloidal springs discussed above, including the chiral nature of particle-induced distortions and defects, the diffusion anisotropy and the presence of two branches of $F_{LdG}(\theta)$. However, these particles preferentially align with their axes nearly parallel to $\mathbf{n}_0$ (Fig. 3d-f,h), which corresponds to the global minimum of $F_{LdG}(\theta)$ for them (Fig. 3i), whereas the nearly orthogonal orientations (Fig. 3a-c,g) now correspond to metastable states. These helices exhibit even stronger dependence of diffusion anisotropy on the particle orientation than their colloidal spring counterparts (Fig. 3g,h). The small-$\theta$ branch of the free energy corresponds to $\mathbf{n}(\mathbf{r})$-configurations with two $s=1$ boojums located at the edges of the end faces, while the large-$\theta$ branch features several additional self-compensating $s=\pm 1$ defects located at the lateral (side) surface of the helix. However, due to the slenderness of the helices, contrary to the case of the springs, the defects nucleation does not promote the strong reduction of the distortions of $\mathbf{n}(\mathbf{r})$, reversing the locations of global and local minima of $F_{LdG}(\theta)$. The dependence of $F_{LdG}(\theta)$ on the direction of tilting is more pronounced for helices (upper left inset in Fig. 3i) than for springs (lower left inset in Fig. 2a), which is caused by the increased ratio of the areas of the end faces to the lateral surface. For tilting of $\hat{\mathbf{e}}_\parallel$ about $\hat{\mathbf{e}}_\perp$ (as in Fig. 2d,e), the large-$\theta$ branches of $F_{LdG}(\theta)$ reveal no local minima (Fig. 3i) because the misalignment of the end faces with respect to $\mathbf{n}_0$ is energetically costly. The comparison of particle-induced director distortions by the colloidal springs and helices shows how this qualitative change of equilibrium orientation emerges from the minimization of distortions (Fig. 2f-i and Fig. 3j-m), with the experimentally observed equilibrium orientations well explained by the minimization of $F_{LdG}(\theta)$.

**Elastic interactions between chiral particles**

Using laser tweezers and video microscopy, we probe elastic pair interactions between chiral springs, elastically aligned with their $\hat{\mathbf{e}}_\parallel$ axes nearly parallel to $\mathbf{n}_0$, when starting from well-defined initial conditions (Fig. 4 and Supplementary Fig. 4). The like-handed colloidal springs attract when their center-to-center separation vector $\vec{d} \parallel \mathbf{n}_0$, but repel when $\vec{d} \perp \mathbf{n}_0$ (Fig. 4a-c). However, the elastic colloidal forces reverse directions as we flip the handedness of one of the two particles undergoing pair interactions (Fig. 4d-f). In addition to these initial conditions ($\vec{d} \parallel \mathbf{n}_0$ and $\vec{d} \perp \mathbf{n}_0$; Fig. 4), colloidal pair interactions at $\vec{d}$ tilted with respect to $\mathbf{n}_0$ also exhibit a different character of force directionality for like-handed and oppositely-handed springs (Supplementary Fig. 4). The angular dependencies of interaction forces and the time dependencies of $\vec{d}$ (Fig. 4b,e), as well as the interaction forces (insets of Fig. 4b,e), are consistent with what is expected for chiral elastic dipoles recently introduced within the approach of nematostatics[17], where left- and right-handed springs have opposite orientations of the dipole moments. Numerical modeling reproduces details of the experimentally observed colloidal behavior (Fig. 5 and Supplementary Fig. 5) providing additional insights into the physical underpinnings. Thus, in Fig. 5a,b we plot $F_{LdG}$ as a function of the angle $\Psi$ between $\vec{d}$ and $\mathbf{n}_0$ for like- and oppositely-handed springs, respectively, and at a fixed center-to-center distance $d$. The particles' $\hat{\mathbf{e}}_\parallel$ axes are aligned along $\mathbf{n}_0$ and different curves correspond to different orientations of $\vec{d}$ relative to particle frames. In agreement with the experiments and nematostatics[17], the like-handed pair of particles tends to align $\vec{d}$ along $\mathbf{n}_0$ with $F_{LdG}(\Psi)$ having minima at $\Psi \approx 0°$ and $180°$ (Fig. 5a), while the opposite-handed ones align $\vec{d}$ orthogonal to $\mathbf{n}_0$, with $F_{LdG}(\Psi)$ minimized at $\Psi \approx 90°$ (Fig. 5b). The behavior of the free energy versus $d$ at $\Psi=0°$, depicted in Figs. 5c and 5d for pair interactions of like- and opposite-handed springs, respectively, also agrees with the experiments and nematostatics: the like-handed springs attract (Fig. 5c) while opposite-handed ones repel (Fig. 5d) along $\mathbf{n}_0$. At relatively large inter-particle distances (see experimental data in Fig. 4b,e and numerically computed dependencies in Fig. 5c,d), colloidal interaction potentials and forces scale with distance as $\propto d^{-3}$ and $\propto d^{-4}$, respectively, consistent with the dipolar nature of interactions. At small distances, the departures from these power law dependencies are caused by the influence of the higher-order elastic multipoles and nonlinear near-field effects that cannot be captured within the multipole expansion analysis, but are revealed by numerical modeling. For example, $F_{LdG}$ exhibits a local minimum at small distances (Fig. 5d), which is related to boojum sharing between proximal end faces.

Both experimental and computational studies of pair interactions reveal the well-defined role of chirality in controlling colloidal elastic pair interactions, which can be generalized to other chiral colloids in LCs and used in conjunction with particle shape aspects to pre-define self-assembly. To provide an example of how the interplay of shape and chirality can further enrich colloidal behavior, we probe pair interactions between colloidal helices (Supplementary Fig. 6) and elucidate many subtle effects. The colloidal helices are found to attract starting from initial conditions with all possible orientations of $\vec{d}$ relative to $\mathbf{n}_0$ and for both like- and opposite-handed pairs of colloidal helices. Videomicroscopy reveals that particles rotate around their $\hat{\mathbf{e}}_\parallel$ axes while undergoing these interactions. The combination of rotational and translational motion yields colloidal assemblies different from the assemblies of colloidal springs (compare Fig. 4 and Supplementary Fig. 6). The nematic colloidal behavior of both colloidal springs and helices can be understood based on nematostatics[17], where one can think about the chiral elastic dipoles being accompanied by additional non-chiral dipole moments arising from the finite dimensions of particles (and the boundary conditions for the director at their end faces)[17], which, in general, can be along $\mathbf{n}_0$ or orthogonal to it. Nematostatics distinguishes four "pure" elastic dipole types: isotropic, anisotropic, chiral and longitudinal (along $\mathbf{n}_0$) ones, each characterized by the corresponding strength parameter (see Supplementary Information for a brief description[17]). Depending on the symmetry of the particle-induced $\mathbf{n}(\mathbf{r})$, its dipolar component can be of a mixed type when more than one dipole strength parameters are different from zero. Since 180°-rotation around $\hat{\mathbf{e}}_\perp$ transforms the finite-length colloidal springs and helices (with director distortions) into themselves (Figs. 1 and 3), these nematic colloidal particles can be classified as general dipoles with $C_2$ point group symmetry, which allows all four pure dipole types, with their strength parameters depending on the details of shape[17]. Both springs and helices have strongly pronounced chiral dipoles, but the overall colloidal behavior is further altered by details of their geometric shape that could be modeled as superposition of effects due to all different types of allowed nematostatic dipoles[17]. For example, unusual attractive interactions between pairs of colloidal helices arise from rotations of these particles around $\mathbf{n}_0$, which lead to the mutual alignment of non-chiral dipole moments orthogonal to $\mathbf{n}_0$ of two chiral helices so that the interactions are attractive. Computational modeling of nematic colloids with different geometry (Supplementary Figs. 7-9) supports this observation and provides additional insights into this behavior. Supplementary Fig. 7 presents the free energy cost $F_{LdG}$ of elastic distortions at fixed $d$ and for $\vec{d} \perp \mathbf{n}_0$, which are induced by the rotation of the springs about their axis $\hat{\mathbf{e}}_\parallel$ in opposite directions by equal amounts. The free energy versus the relative angle $\varphi_2-\varphi_1$ between particles $\hat{\mathbf{e}}_\perp$ axes has two minima of equal potential depths at $\varphi_2-\varphi_1=0°$ and 360°, which are separated by a large barrier. To minimize free energy, particles tend to align their $\hat{e}_\perp$ axes parallel to each other. The functional form of $F_{LdG}(\varphi_2-\varphi_1)$ is well approximated by the expression $A+B\cos(\varphi_2-\varphi_1)$, as expected for particles whose elastic dipole is a mixture of the chiral and the longitudinal types[17]; parameters $A$ and $B$ are related to the corresponding dipolar strengths. The barrier separating two minima of $F_{LdG}$ for like-handed springs is much higher than for the opposite-handed ones (Supplementary Fig. 7). This analysis, along with the insights provided by nematostatics[17], supports the notion that the effects of chirality and geometric shapes can be effectively coupled to achieve better control of colloidal self-assembly, as also evident from the comparison of alignment and self-assembly of colloidal springs and helices (Figs. 1-5 and Supplementary Figs. 4-7).

**Colloidal superstructures**

Our analysis of pair and many-body interactions shows that the equilibrium assembly of like-handed helices with the axes $\hat{e}_\parallel$ roughly along $n_0$ is also chiral in nature, with the equilibrium orientation of $\vec{d}$ within the assembly along $n_0$ (Fig. 5e). In contrast, like-handed springs form chains of particles with $\vec{d} \parallel \hat{e}_\parallel$ and both orthogonal to $n_0$ (Fig. 5f). This differs from the racemic dispersions of oppositely-handed helices and springs, which both form equilibrium assemblies in the form of chains with $\vec{d} \perp n_0$ (Fig. 5g,h). The $\hat{e}_\parallel$ axes of helices are, on average, roughly parallel to $n_0$ (Fig. 5g), though their slight tilting direction alternates between oppositely-handed particles within the chain. The $\hat{e}_\parallel$ axes of springs are roughly orthogonal to $n_0$ (Fig. 5h), albeit their tilt direction also alternates within the chains. The main mechanism of the synclinic particle tilting (Fig. 5e,f) is due to the single particle's tendency to equilibrate its axes at some small angles with respect to $n_0$ (Figs. 1c,e-j, 2a,f and 3d-f,h) or the plane orthogonal to $n_0$ (Figs. 1d,k-n, 2a,h and 3a-c,g). In assemblies of particles with opposite handedness (Fig. 5g,h), the anticlinic tilting of $\hat{\mathbf{e}}_{\|}$ away from $\mathbf{n}_0$ arises from the pairwise LC-induced torques (Fig. 6 and Supplementary Figs. 8a and 9). Although these self-assembled configurations exhaust all possible energy-minimizing colloidal structures in the bulk of LC, confinement can further enrich colloidal interactions between chiral particles. For example, cell or capillary confinement can preclude bulk-type ordering of like-handed springs and helices (Fig. 5e) by enforcing $\vec{d}$ to be orthogonal to $\mathbf{n}_0$. At high number densities of particles, this confinement can lead to chiral colloidal configurations emerging from repulsive-only interactions between particles (Supplementary Fig. 9b).

We compute $F_{LdG}$ versus the relative alignment of two like-handed springs and $d$ at $\vec{d} \perp \mathbf{n}_0$ (Fig. 6a,b). When $\hat{\mathbf{e}}_{\|}$ axes rotate about $\hat{\mathbf{e}}_{\perp}$ (Fig. 6e), the free energy as a function of the angle $\Omega_\perp$ between the $\hat{\mathbf{e}}_{\|}$ axes has a local minimum at $\Omega_\perp \approx 16°$ (open circles in Fig. 6c) and a global minimum at $\Omega_\perp \approx 176°$ (open circles in Fig. 6d), indicating the intrinsic tendency of springs to twist when they are placed side-by-side, with $\vec{d} \perp \mathbf{n}_0$. Supplementary Fig. 8a illustrates the twisting tendency of like-handed helices: $F_{LdG}(\Omega_\perp)$ shows only global minimum at $\Omega_\perp = 12°$. Contrary to the case of the springs, this configuration for helices is stable relative to the variation of $d$ (Supplementary Fig. 8c), resulting in helicoidal assemblies of particles. Opposite-handed springs aligned along $\mathbf{n}_0$ also exhibit the twisting tendency when placed side-by-side (see the data presented by squares in Fig. 6c,d). However, the twisting direction is sensitive to the placement of springs relative to each other (compare the open and the solid squares in Fig. 6c,d and Supplementary Fig. 9a). Racemic dispersions of left- and right-handed chiral particles do not possess helicoidal ordering (Fig. 5g,h), showing that opposite-handed chiral particles cannot form chiral superstructures.

**Conclusions**

We have demonstrated that chirality of colloidal particles interplays with the nematic elasticity to pre-define chiral or racemic superstructures. Chirality induced by dilute colloidal dispersions competes with the boundary conditions on confining surfaces to yield unwound configurations in which the LC host has a uniform far-field director prescribed by boundary conditions on confining substrates, but the colloidal superstructures can be either chiral helicoidal or racemic, depending on the relative chirality (same or opposite) of the suspended particles. Our findings demonstrate a means of pre-determining self-assembly by controlling handedness of individual colloidal constituents, which interplays with the particle's geometric shape[20,21] to provide a rich framework for designing and realizing complex chiral nematic colloidal composites. These findings may lead to the realization of enantiomorphic colloidal crystals and chiral molecular-colloidal hybrid LC fluids with properties controlled through self-assembly at the mesoscale. The frustration resulting from the competition between the chiral nature of the colloidal superstructures and the uniform far-field background defined by the strong surface boundary conditions can result in a host of localized field configurations, such as solitons, which were recently extensively studied[34] in frustrated chiral nematic LCs with molecular chiral additives.

## Methods

**Preparation of chiral nematic colloids.** Colloidal microparticles in the forms of springs and helices of different handedness were fabricated using a home-built two-photon photopolymerization setup shown in Supplementary Fig. 2. Each particle was obtained by advancing the focus of a femtosecond laser beam along a computer-programmed three-dimensional trajectory mimicking the desired shape of a chiral particle. When this laser beam's focus moved within the bulk of the photoinitiator-containing commercial monomeric mixture IP-L (obtained from NanoScribe GmbH), this resulted in the photopolymerization of a solid tube forming a chiral particle with pre-defined geometric parameters. By prescribing a spring trajectory with different parameters, we generate surface-attached polymerized microstructures shaped as a spring or a helix. The chiral particle's tube diameter was varied from 0.3 to 3μm while the overall size of particles was controlled within the range from 3 to 15μm (Fig. 1 and Supplementary Figs 1,2). Once detached from surfaces of substrates using mechanical agitation, these colloidal particles were re-dispersed in an isotropic solvent (isopropanol). Following this, the chiral colloidal objects were transferred into either a single-compound nematic LC pentylcyanobiphenyl (5CB) or into a room-temperature ZLI-2806 nematic mixture (both obtained from EM Chemicals). Colloidal springs and helices induced strong tangentially degenerate surface boundary conditions without further surface treatment or chemical functionalization. Using capillary forces, nematic colloidal dispersions of springs and helices were infiltrated into glass cells with the gap thickness ranging within 20-60 μm, which were made of two parallel glass plates with inner surfaces treated to induce strong planar or perpendicular surface boundary conditions for the LC director field $\mathbf{n}(\mathbf{r})$[6,7,35]. The details of the surface treatment that allows one to set these boundary conditions are described elsewhere[35-38].

**Three-dimensional optical imaging and laser trapping.** In our study, we utilize an integrated multi-functional optical setup capable of simultaneous 3PEF-PM imaging, holographic optical trapping (HOT, operating at 1,064 nm), and also conventional optical bright-field imaging and transmission-mode polarizing microscopy. This multi-modal imaging and manipulation setup was built around an inverted optical microscope IX81 (purchased from Olympus)[7]. For the 3PEF-PM imaging, we have employed a tunable (within the spectral range of 680-1080 nm) Ti-Sapphire femtosecond oscillator (Chameleon Ultra II, Coherent) as the laser excitation source, which is emitting 140 fs pulses at a repetition rate of 80 MHz. The chiral colloidal particles and the chiral director structures in the nematic LC host around them were imaged by utilizing the polarization-dependent fluorescence signals arising due to the nonlinear optical processes involving multi-photon absorption of LC molecules within the dispersion of chiral colloids in a nematic host[36,37]. To optimize the imaging conditions, the excitation wavelength of the femtosecond laser source was tuned to 870 nm, yielding strong three-photon absorption-based polarized excitation of the 5CB molecules. The ensuing 3PEF-PM signals were collected in the epi-detection mode. A commercially available photomultiplier tube (H5784-20, Hamamatsu) was used as a detector. An Olympus 100× oil-immersion objective was used for both optical imaging and laser trapping[6,38], where its high numerical aperture of 1.4 allowed us to obtain good optical resolution.[35-38]

This type of imaging enabled the visualization of particle orientation within the LC on the basis of the contrast of fluorescence intensity. Moreover, unlike in the case of the (isotropic) IP-L polymer inside particles, the fluorescence intensity that arises from the nematic host 5CB molecules depends on the orientation of the linear polarization direction set by a polarizer within the setup's excitation channel (Fig. 1l-n and Supplementary Fig. 2). The three-dimensional tables containing fluorescence intensity data dependent on the spatial coordinates within the sample and on the polarization states of the excitation light are stored in a computer and then used to experimentally reconstruct the director field configurations. The analysis of such polarization-dependent 3PEF-PM image stacks composed of individual optical "slices", such as the ones presented in Fig. 1l-n, reveals the dependence of the particle-induced **n(r)** and topological defects on the geometric shape of our chiral colloidal particles in the nematic fluid host.

**Optical video microscopy and particle tracking.** Photopolymerized individual chiral colloidal particles stay suspended in the bulk of a nematic LC host fluid. This colloidal stability is facilitated by their Brownian motion due to thermal fluctuations. Furthermore, elastic repulsions of the chiral particles from the confining substrates of the glass cells also balance the gravitational forces and help precluding their sedimentation. To gain insights into the diffusion of particles in the nematic fluid, we used bright field optical microscopy and video tracking. We determine the lateral position of a particle within each frame of a video using freely available software (ImageJ and its plugins, from the National Institute of Health). We then analyze the translational displacements of chiral colloidal particles with the regular time steps corresponding to the frame rates of the video. This allows us to build the histograms of particle displacements, such as the ones shown in Figs. 1c,d and 3g,h. We fit obtained experimental histograms with Gaussian distributions and experimentally determine two independent diffusion coefficients[16,21] $D_\parallel$ and $D_\perp$, which characterize the diffusion of particles along and perpendicular to $\mathbf{n}_0$, respectively. The Stokes-Einstein relation then also allows us to determine the corresponding direction-dependent friction coefficients on the basis of experiments. Subsequently, the estimated friction coefficients are used for determining the anisotropic colloidal interaction forces between our chiral particles[6,16,21]. In the video microscopy experiments, we use a CCD camera (Flea, from PointGrey) to record videos of motion of colloidal particles at the frame rate of 15 frames per second. The lateral positions of chiral particles versus time are determined from captured video sequences with the spatial precision[39] in the range of 7-10 nm.

**Methods and procedures of numerical modeling.** The interplay of chirality of the chiral colloidal particle shape and **n(r)** is also explored by a numerical modeling approach. Our numerical procedure is based on the minimization of the Landau-de Gennes free energy[1,25]. The free energy functional of the LC, expressed in terms of the order parameter tensor **Q**, combines the nematic elasticity, the variable nematic degree of order, and the surface anchoring terms[1,25]. The numerical minimization of the total free energy yields both the theoretical characterization of **n(r)** orientation and the local changes in the nematic degree of order that correspond to global or local minima of the free energy[25]. The colloidal particle surfaces are defined as described below, with the parameters tuned to match the geometric features of the corresponding experimental counterparts. Minimization of the total free energy is performed numerically for the tangentially degenerate surface boundary conditions using a variable three-dimensional grid[7,22]. This minimization yields stable or metastable **n(r)** structures around particles, which are then directly compared to the experimentally reconstructed counterparts[22].

Nematic director configurations around colloidal helices and springs are obtained via numerical minimization of the phenomenological Landau-de Gennes free energy functional[25]

$$F_{LdG} = \int_V \left( a Q_{ij}^2 - b Q_{ij} Q_{jk} Q_{ki} + c \left( Q_{ij}^2 \right)^2 + \tfrac{L_1}{2} \partial_k Q_{ij} \partial_k Q_{ij} + \tfrac{L_2}{2} \partial_j Q_{ij} \partial_k Q_{ik} \right) dV + W \int_{\partial V} f_s(Q_{ij}) dS, \quad (1)$$

where $Q_{ij} = Q_{ji}$ $(i, j = 1,...,3)$ is a traceless tensor order parameter and summation over repeated indices is assumed. In equation (1), the parameter $a$ (unlike the constants $b$ and $c$) is assumed to depend linearly on temperature $T$: $a(T)=a_0(T-T^*)$, where $a_0$ is a material dependent constant, and $T^*$ is the supercooling temperature of the isotropic phase. Phenomenological parameters $L_1$ and $L_2$ are related (via an uniaxial Ansatz for $Q_{ij}$) to the Frank-Oseen elastic constants. We describe planar degenerate anchoring, with the strength coefficient $W$, of colloidal particles by using

$$f_s(Q_{ij}) = (\tilde{Q}_{ij} - \tilde{Q}_{ij}^{\perp})^2 + (\tilde{Q}_{ij}^2 - 3Q_b^2/2)^2 \text{ with } \tilde{Q}_{ij} = Q_{ij} + Q_b \frac{\delta_{ij}}{2}, \tilde{Q}_{ij}^{\perp} = (\delta_{il} - v_i v_l)\tilde{Q}_{lk}(\delta_{kj} - v_k v_j) \text{ and with } \delta_{ij}$$

being the Kronecker delta symbol, and $v$ is the unit outward vector normal to the confining surface[40]; $Q_b = b/8c\left(a + \sqrt{1 - 64ac/(3b^2)}\right)$ is the value of the scalar order parameter in the nematic phase, which is thermodynamically favored for $24ac/b^2 < 1$. Colloidal particle surfaces are defined to match the geometry of the corresponding experiments. Minimization of the free energy [equation (1)] is then performed numerically by employing adaptive mesh finite elements method as described in more details in Ref. 41. This minimization yields stable or metastable $\mathbf{n}(\mathbf{r})$ configurations around particles. In our calculations, we use $a_0 = 0.044 \times 10^6$ J/m$^3$, $b = 0.816 \times 10^6$ J/m$^3$, $c = 0.45 \times 10^6$ J/m$^3$, $L_1 = 6 \times 10^{-12}$ J/m, and $L_2 = 12 \times 10^{-12}$ J/m, which are typical values for 5CB[42] at $T^* = 307$ K. For these values of the model parameters, the bulk correlation length $\xi = 2\sqrt{2c(3L_1 + 2L_2)}/b \approx 15$ nm at the isotropic-nematic coexistence[43], $24ac/b^2 = 1$.

**Geometry of colloidal helices and springs.** To define colloidal helices and springs using two-photon photo-polymerization, we exploit the following parametrization in Cartesian coordinates:

$$\mathbf{r}(\tau) = \begin{pmatrix} x(\tau) \\ y(\tau) \\ z(\tau) \end{pmatrix} = \begin{pmatrix} R\cos\tau \\ \pm R\sin\tau \\ h\tau/4\pi \end{pmatrix} \quad \tau \in [0, 4\pi], \quad (2)$$

which defines a circular right-handed (+) or left-handed (-) helices and springs of radius $R$ and the helical pitch $h/2$. In our two-photon photo-polymerization experiments, we set $h = 12$ μm and use $R$=6 μm for springs and $R$=3 μm for helices. In the theoretical modeling, we exploit the Open Source Gmsh library[44] to triangulate the surface of a helix. We set $h$=1.25 μm, and $R$=0.5 μm and $R$=0.1 μm for the spring and single helix colloids, respectively. In both cases particles have a circular cross-section of radius $r$=0.1 μm. Then the center of mass $\mathbf{r}_c$ of a colloidal particle generated by equation (2) is defined as $\mathbf{r}_c$=(0,0,$h$/2)$^T$, where "T" stands for "transpose." To describe configurations of chiral colloidal particles relative to the far field director $\mathbf{n}_0$ or with respect to each other, we rigidly attach a right-handed orthonormal reference frame ($\hat{\mathbf{e}}_\parallel$, $\hat{\mathbf{e}}_\perp$, $\hat{\mathbf{e}}_3$) to each particle as follows (Supplementary Fig. 1): the first unit vector $\hat{\mathbf{e}}_\parallel$ is parallel to the particle long axis and is oriented in the direction of increasing $\tau$, the second unit vector $\hat{\mathbf{e}}_\perp$ is perpendicular to the particle axis and is oriented towards the spring/helix beginning point (corresponding to $\tau$=0); the third vector is $\hat{\mathbf{e}}_3$=$\hat{\mathbf{e}}_\parallel \times \hat{\mathbf{e}}_\perp$. For example, in the coordinate system corresponding to equation (2), we have $\hat{\mathbf{e}}_\parallel$=$\hat{\mathbf{e}}_z$, $\hat{\mathbf{e}}_\perp$=$\hat{\mathbf{e}}_x$, $\hat{\mathbf{e}}_3$=$\hat{\mathbf{e}}_y$.

**Code availability.** All codes used in this work are freely available from the corresponding author upon a request.

**Data availability.** The data that support the findings of this study are available from the corresponding author on reasonable request.

**Acknowledgements**

We thank P. Davidson and T. Lubensky for discussions. We acknowledge support of the U.S. Department of Energy, Office of Basic Energy Sciences, Division of Materials Sciences and Engineering, under Award ER46921, contract DE-SC0010305 with the University of Colorado

Boulder, as well as partial support of the American Chemical Society Petroleum Research Fund Grant PRF 54095-ND7 (development of the instrument for particle fabrication).

**Author Contributions**

Y.Y., A.M., B.S., and I.I.S. conducted experimental work and analyzed data. M.T. performed numerical modeling. M.T. and I.I.S. wrote the manuscript, with the input from all authors. I.I.S. conceived and designed the project.

**Additional information**

Supplementary information is available in the online version of the paper. Reprints and permissions information is available at www.nature.com/reprints. Correspondence and requests for materials should be addressed to I.I.S.

**Competing financial interests**

The authors declare no competing financial interests.

**Figures and captions**

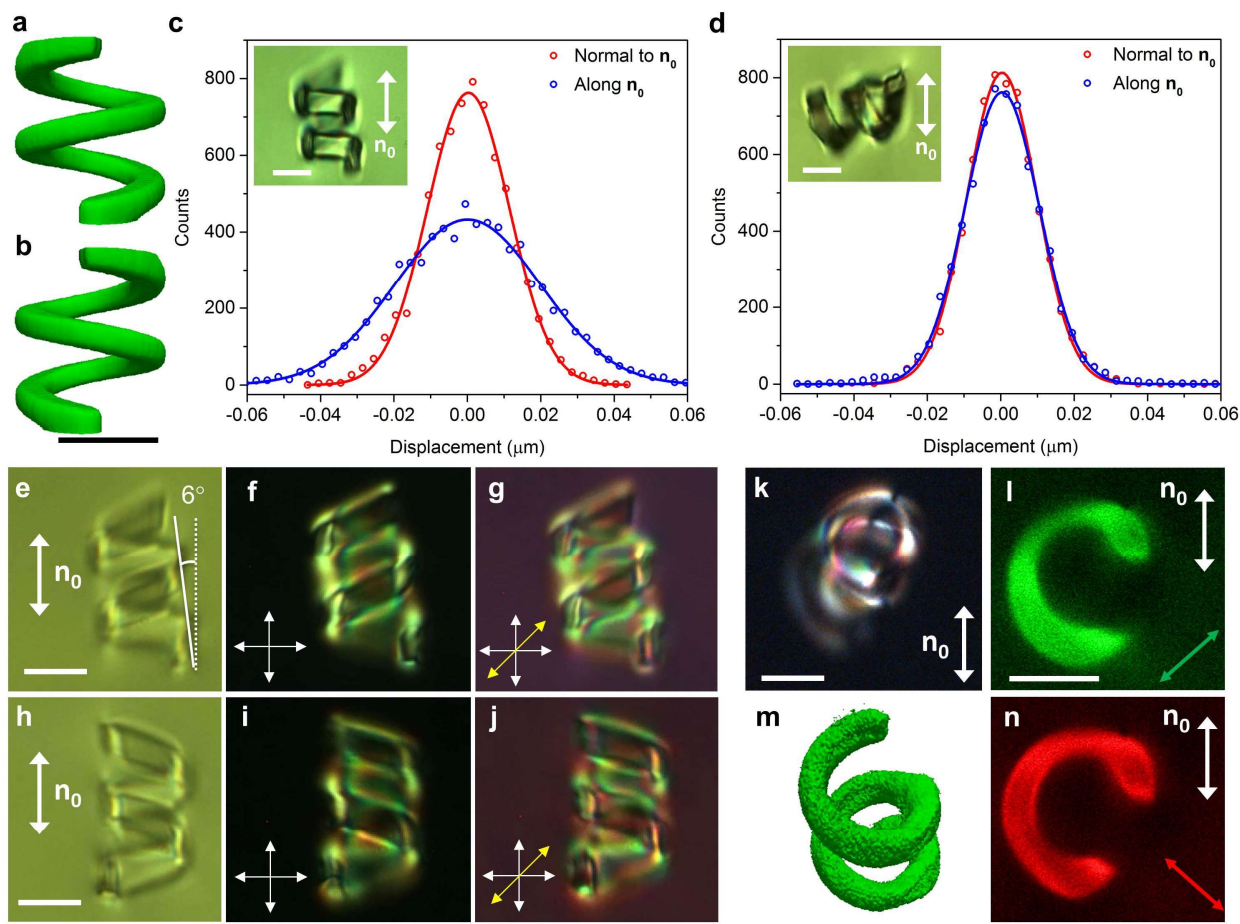

**Figure 1 | Microsprings in a nematic liquid crystal. a**, **b**, Right- and left-handed colloidal springs imaged with the help of 3PEF-PM while surrounded by an isotropic medium (immersion oil). **c,d**, Self-diffusion of the particles oriented with the spring axis roughly **c**, parallel and **d**, perpendicular to $\mathbf{n}_0$. The insets show optical bright field micrographs at the corresponding orientations with respect to $\mathbf{n}_0$. Solid lines are Gaussian fits to experimental data allowing to calculate diffusion coefficients[12] along and normal to $\mathbf{n}_0$ (**c**, $D_\parallel=2.9\times10^{-3}$ μm$^2$/s, $D_\perp=0.93\times10^{-3}$ μm$^2$/s and **d**, $D_\parallel=3.0\times10^{-3}$ μm$^2$/s, $D_\perp=2.7\times10^{-3}$ μm$^2$/s). **e-j**, Optical micrographs of **e-g**, right and **h-j**, left-handed colloidal microsprings in 5CB imaged using **e**, **h**, bright field and polarizing microscopy **f**, **i**,

without and **g**, **j**, with an additional phase retardation plate (with the slow axis marked by the yellow double arrow) inserted between the crossed polarizers (white double arrows). **k-n**, Detailed analysis of **n(r)**-distortions induced in the LC by colloidal springs probed for the same particle using **k**, polarizing microscopy and **l-n**, 3PEF-PM, with **l** and **n** showing details of two individual in-plane 3PEF-PM optical slices obtained for two different linear 3PEF-PM polarizations (marked by the green and red double arrows, respectively) and **m** showing the corresponding 3D perspective view of the particle reconstructed on the basis of superposition of many such slices. Scale bars are 5 μm.

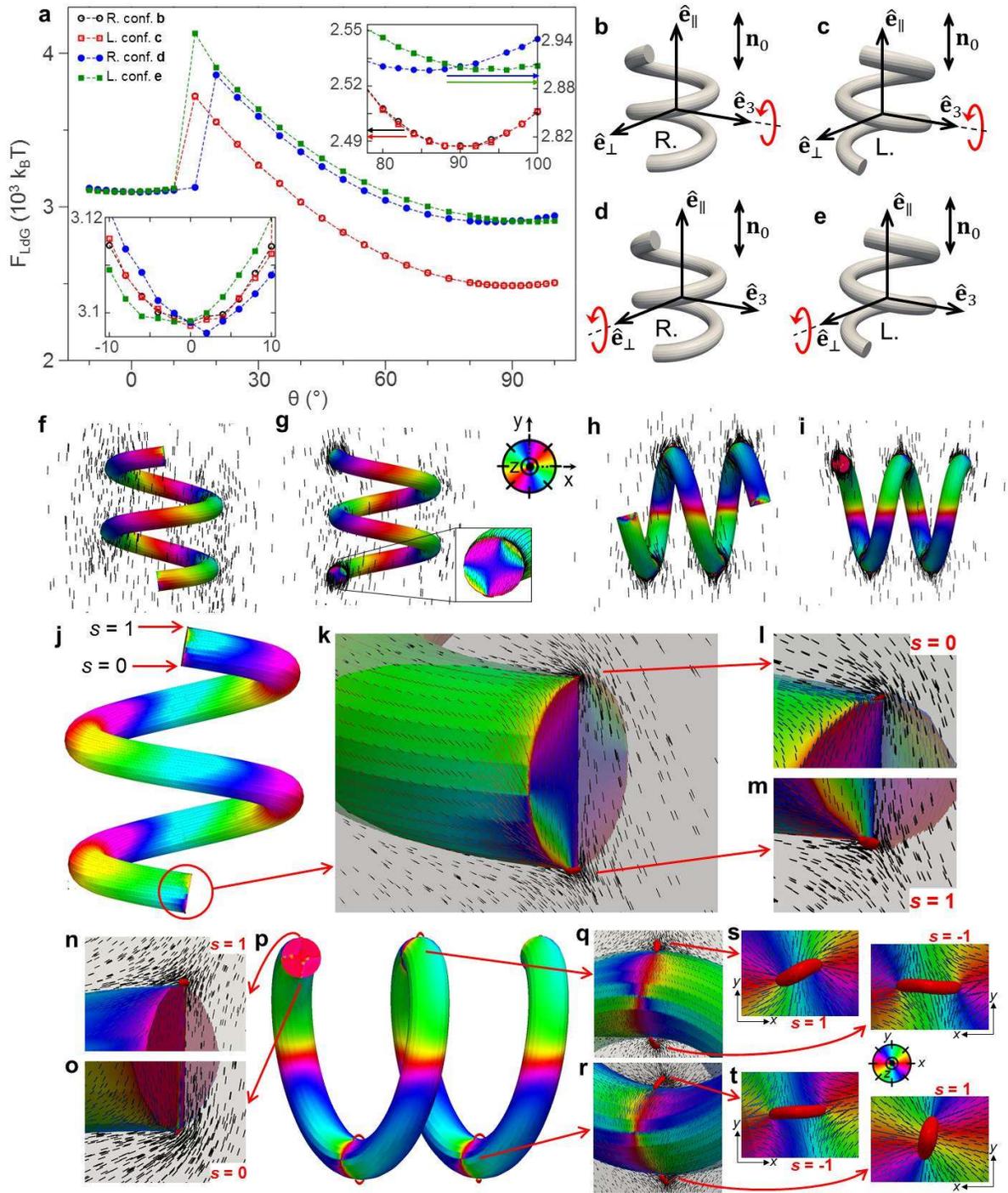

**Figure 2 | Chirality-dictated alignment of microsprings in a nematic liquid crystal. a**, Orientation-dependent free energy cost of particle-induced elastic distortions versus the angle between the spring axis $\hat{\mathbf{e}}_{\parallel}$ (see **b-e** and Supplementary Fig. 1 for the definition of the orthonormal

reference frame attached to the particle) and $\mathbf{n}_0$ for different tilting direction of $\hat{\mathbf{e}}_\parallel$, and for the right- (circles) and left-handed (squares) particles. Open (solid) symbols correspond to the anticlockwise rotation of $\hat{\mathbf{e}}_\parallel$ about $\hat{\mathbf{e}}_3(\hat{\mathbf{e}}_\perp)$ as shown in **b-e**. The insets depict free energy variations near the stable and metastable orientation states. **f-i**, director structures around right-handed springs at energy-minimizing orientations with **f,** $\theta=4°$ (blue solid circles in **a**) and with **h,** $\theta=86°$ (blue solid circles in **a**), and left-handed springs with **g,** $\theta=0°$ (red open squares in **a**) and with **i,** $\theta=90°$ (red open squares in **a**) for the **f,g**, metastable and **h,i**, stable particle orientations. **j-m**, Detailed director structures and defects corresponding to a colloidal spring shown in **g**. **n-t**, Detailed director structures and defects corresponding to a spring shown in **i**. Director structures are shown with rods depicting local orientations of $\mathbf{n}(\mathbf{r})$ both at LC-particle interfaces and in the bulk, as well as with color-coded patterns of azimuthal orientations of $\mathbf{n}(\mathbf{r})$ when projected from particle surfaces to the plane orthogonal to $\mathbf{n}_0$, according to the color scheme in the insets of **g** and **t**, as viewed from different perspectives.

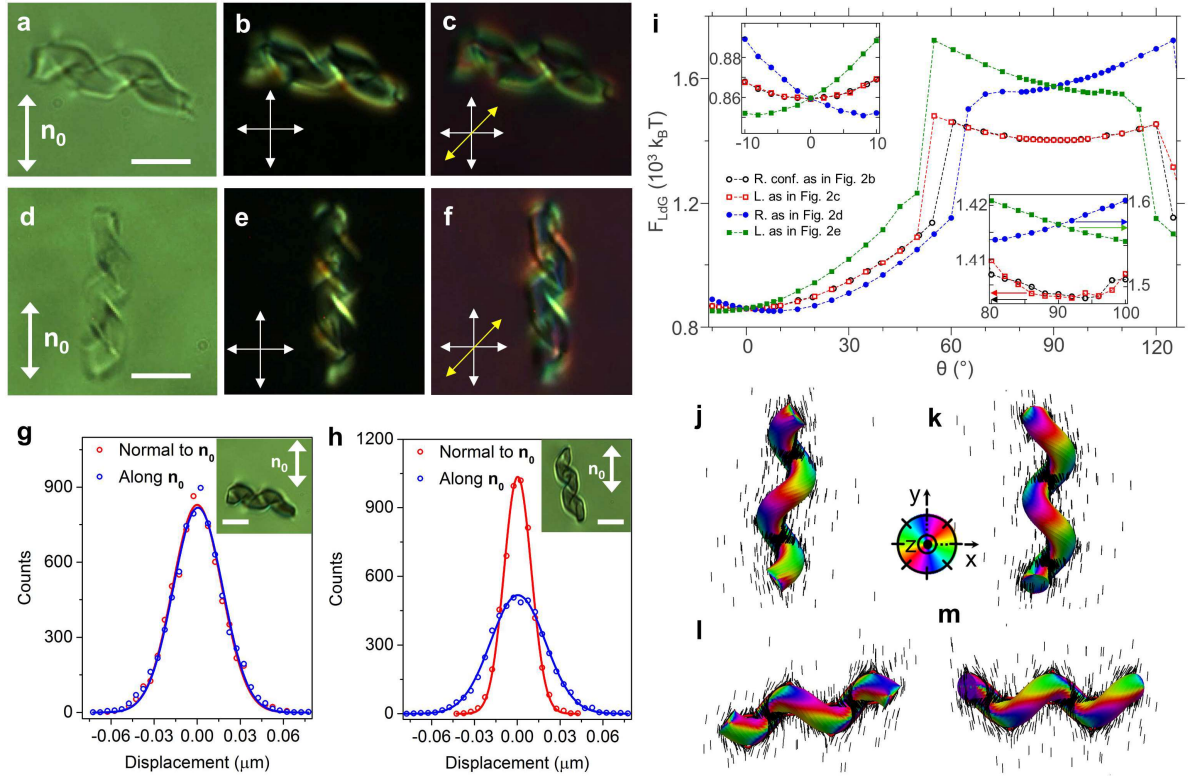

**Figure 3 | Chirality-dictated alignment of single helices in a nematic liquid crystal. a-f**, Optical micrographs of left-handed microhelices obtained using **a**, **d**, bright field and polarizing microscopy **b**, **e**, without and **c**, **f**, with a phase retardation plate. **g**, **h**, Self-diffusion of particles oriented with $\hat{\mathbf{e}}_\parallel$ roughly **g**, orthogonal and **h**, parallel to $\mathbf{n}_0$. The insets show optical bright field micrographs at the corresponding orientations with respect to $\mathbf{n}_0$ (**g**, $D_\parallel \approx D_\perp \approx 2.2 \times 10^{-3}$ μm²/s and **h**, $D_\parallel \approx 2.7 \times 10^{-3}$ μm²/s, $D_\perp \approx 0.68 \times 10^{-3}$ μm²/s). **i**, orientation-dependent free energy cost of elastic distortions induced by particles versus the angle between $\hat{\mathbf{e}}_\parallel$ and $\mathbf{n}_0$ for different orientations of the tilting direction of $\hat{\mathbf{e}}_\parallel$, and for the left- (squares) and right-handed (circles) helices. Open (solid) symbols correspond to the anticlockwise rotation of $\hat{\mathbf{e}}_\parallel$ about $\hat{\mathbf{e}}_3(\hat{\mathbf{e}}_\perp)$ as shown in Fig. 2b-e. The insets depict free energy variations near the stable and metastable orientation states. **j-m**, Director structures around right-handed helices at **j,** $\theta=8°$ (blue solid circles in **i**) and at **l,** $\theta=82°$ (blue solid circles in **i**) and left-handed helices at **k,** $\theta=0°$ (red open squares in **i**) and **m,** $\theta=90°$ (red open squares in **i**), which are **j, k**, stable and **l, m**, metastable particle orientations. The director field is shown with the help of rods and color-coded azimuthal orientations of $\mathbf{n}(\mathbf{r})$ on particle surfaces relative to $\mathbf{n}_0$ according to the color scheme shown in the inset of **k**. Scale bars are 5 μm.

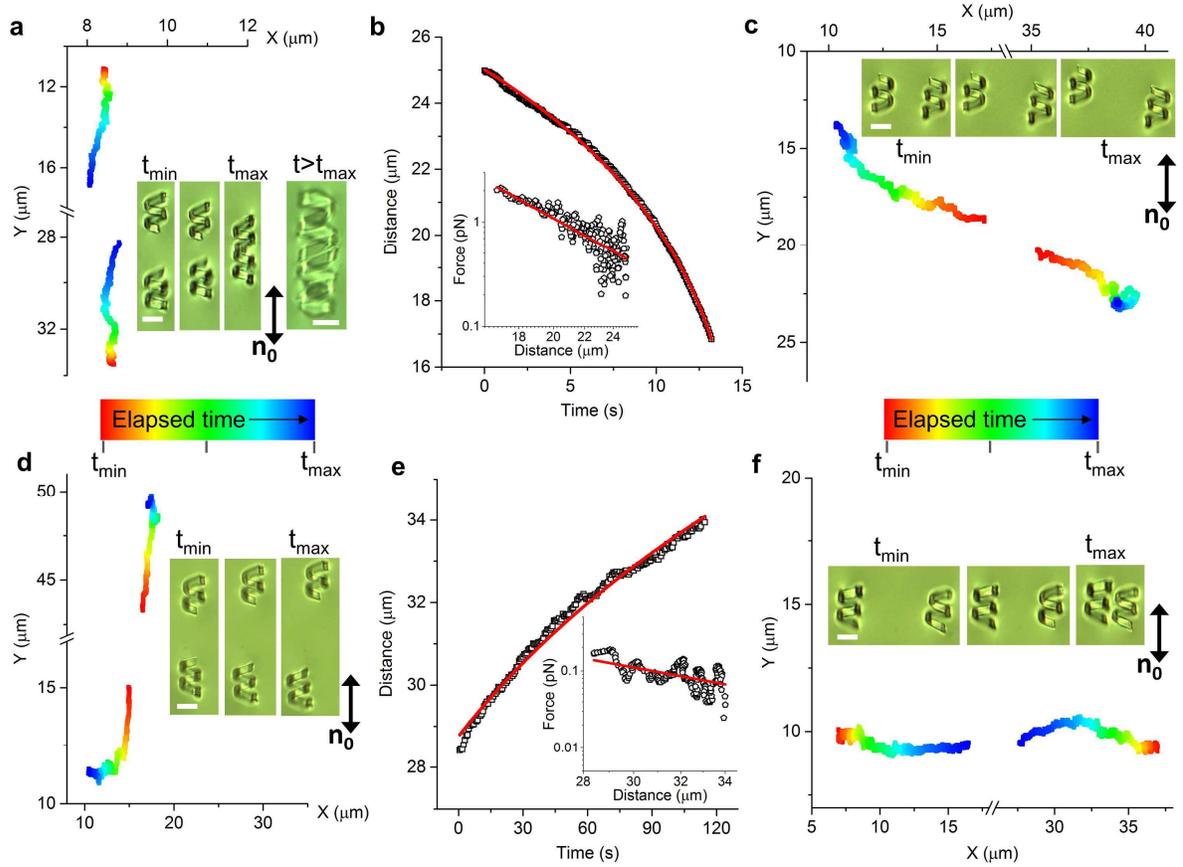

**Figure 4 | Chirality-dependent pair interactions of colloidal springs. a**, Time-color-coded trajectory of attraction of like-handed microsprings initially separated along $\mathbf{n}_0$. The bottom inset shows colors scale of elapsed time counted from the moment of releasing particles from laser traps ($t_{min}$) till the moment when they approach each other at $t_{max}$ ($t_{max}-t_{min}=27s$), as shown in the bright field micrographs in the inset. The final self-assembled colloidal structure for the same particles at $t=35s > t_{max}$ is also shown in the inset in a slightly enlarged micrograph and emerges as a result of additional rotation of the colloidal springs around their axes. **b**, Separation distance $d$ versus time corresponding to **a**, with the distance-dependence of the corresponding interaction force shown in the inset. **c**, Time-color-coded trajectory of repulsion of like-handed microsprings initially separated so that the separation vector is orthogonal to $\mathbf{n}_0$. **d**, Time-color-coded trajectory of repulsion of oppositely-handed microsprings initially separated along $\mathbf{n}_0$. The insets show the corresponding

optical micrographs of the interacting particles. **e**, Separation distance versus time corresponding to **d**; the distance-dependence of the corresponding interaction force is shown in the inset. **f**, Time-coded trajectory of attraction of oppositely-handed microsprings initially separated so that the separation vector is orthogonal to $\mathbf{n}_0$. The red curves in **b**,**e**, are the best fits of the experimental data with $d(t) = (d_0^n - n\alpha t)^{1/n}$ where $n=5$ for dipole-dipole interaction[21]; the fitting coefficients are **b**, $d_0$=25µm, $\alpha$=1.3×10$^5$ µm$^5$/s, and **e**, $d_0$=29µm, $\alpha$=-0.46×10$^5$ µm$^5$/s. Solid red lines in the inset of **b** and **e** are the best linear fits with[17] $\ln F = -(n-1)\ln d + const$. Scale bars are 5 µm.

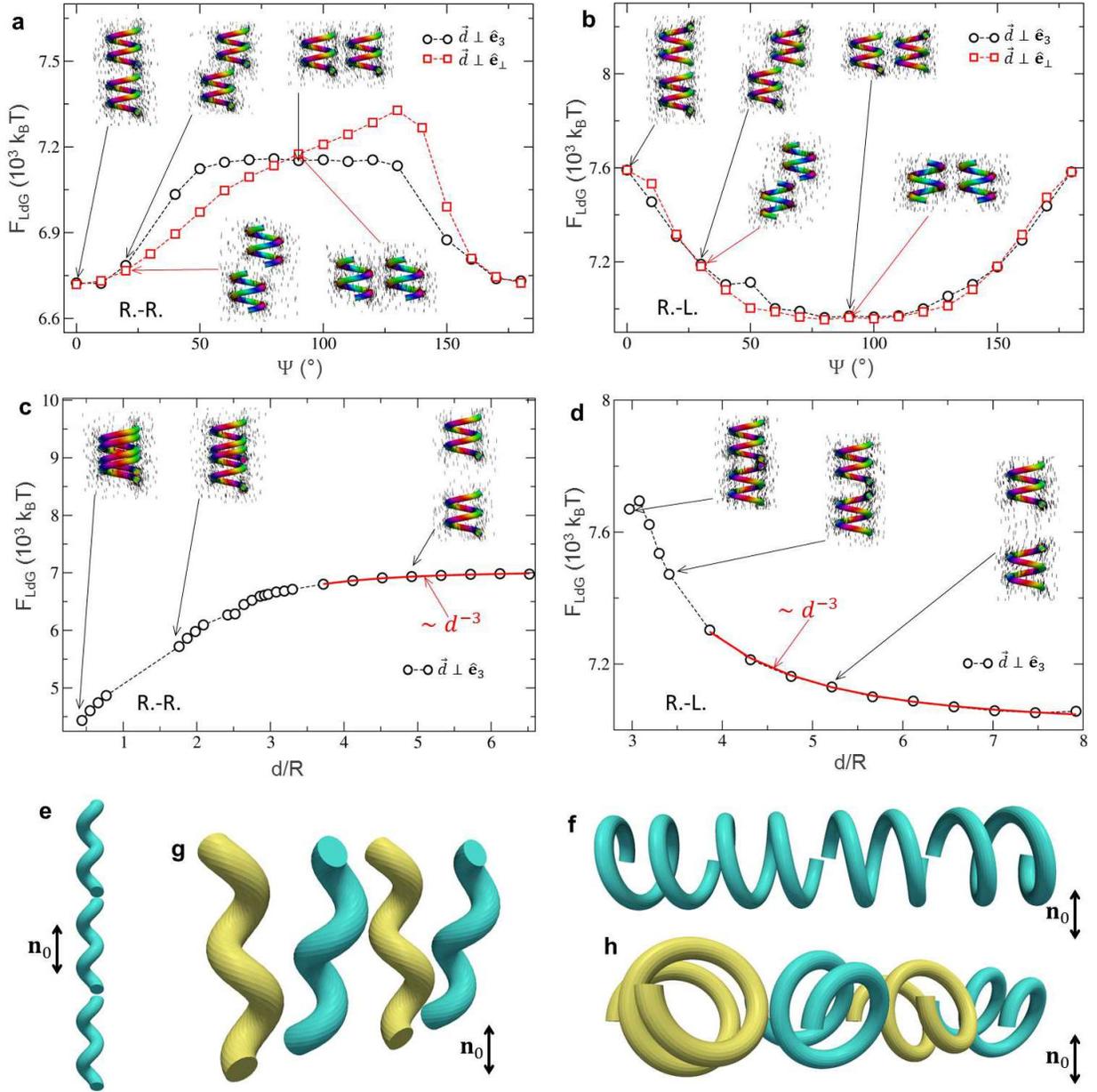

**Figure 5 | Chirality-dependent pair interactions of microsprings. a-d,** Numerically calculated Landau-de Gennes free energy $F_{LdG}$ of a pair of **a**, **c** like- and **b**, **d** oppositely-handed microsprings as **a**, **b** a function of the angle $\Psi$ between the center-to-center vector $\vec{d}$ and $\mathbf{n}_0$ and at $d=1.65R$ (see Methods for definition of $R$), and as **c**, **d** a function of the center-to-center distance $d$ at $\Psi=0°$. The particles' own reference frames are fixed in all cases. Open circles (squares) in **a**, **b** correspond to the case when $\Psi$ varies in the plane spanned by the vectors $\hat{\mathbf{e}}_\parallel$ and $\hat{\mathbf{e}}_\perp$ ($\hat{\mathbf{e}}_\parallel$ and $\hat{\mathbf{e}}_3$). Red curves in **c,d** at

large $d/R$ are the fits of numerical data with the $\propto d^{-3}$ power law expected for the colloidal interaction potential due to dipolar interactions. **e-h**, Bulk equilibrium assemblies of **e**, like-handed and **g**, opposite-handed helices and **f**, like-handed and **h**, opposite-handed springs.

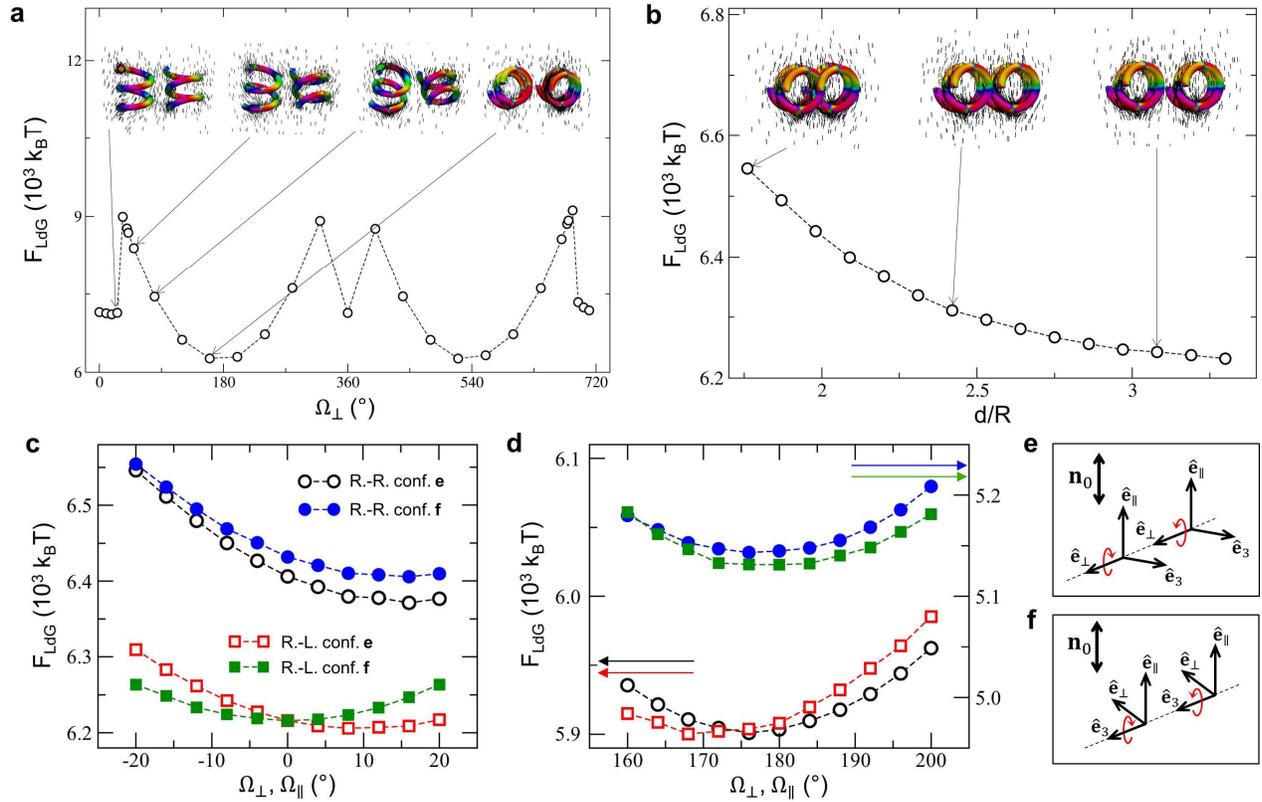

**Figure 6 | Twisting of chiral particles relative to each other. a**, Numerically calculated Landau-de Gennes free energy as a function of the angle $\Omega_\perp$ between the axes of like-handed springs, for the case when these spring axes rotate about the particles $\hat{\mathbf{e}}_\perp$ vectors in opposite directions and by equal amounts, as is shown schematically in a panel **e**. The spring center-to-center vector $\vec{d}$ is fixed with $d=1.625R$ and with the angle $\Psi$ between $\vec{d}$ and $\mathbf{n}_0$ at 90°. **b**, Free energy as a function of $d$ at $\Omega_\perp=180°$, all the other parameters are the same as in **a**. The insets in **a**, **b** show the director configurations around the springs with the help of rods and color-coded azimuthal, with respect to

$\mathbf{n}_0$, orientations of $\mathbf{n}(\mathbf{r})$ on particle surfaces. Panels **c**, **d** represent the case of like- (circles) and opposite-handed (squares) springs, and depict the free energy variations with $\Omega_\perp$ (particles' axes rotate about $\hat{\mathbf{e}}_\perp$) or $\Omega_\parallel$ (particles' axes rotate about $\hat{\mathbf{e}}_3$) at $d=1.625R$ and $\Psi=90°$, near the stable **d** and metastable **c** orientation states. Open (solid) symbols correspond to the case when the spring axes $\hat{\mathbf{e}}_\parallel$ rotate about particles aligned $\hat{\mathbf{e}}_\perp$ vectors, see **e** ($\hat{\mathbf{e}}_3$ vectors, see **f**) in opposite directions by equal amounts.

# Chiral liquid crystal colloids

Ye Yuan,[1] Angel Martinez,[1] Bohdan Senyuk,[1] Mykola Tasinkevych[2,3,4] & Ivan I. Smalyukh[1,5,6*]

[1]*Department of Physics and Soft Materials Research Center, University of Colorado, Boulder, CO 80309, USA*

[2]*Max-Planck-Institut für Intelligente Systeme, Heisenbergstr. 3, D-70569 Stuttgart, Germany*

[3]*IV. Institut für Theoretische Physik, Universität Stuttgart, Pfaffenwaldring 57, D-70569 Stuttgart, Germany*

[4]*Centro de Física Teórica e Computacional, Departamento de Física, Faculdade de Ciências, Universidade de Lisboa, Campo Grande P-1749-016 Lisboa, Portugal*

[5]*Department of Electrical, Computer, and Energy Engineering, Materials Science and Engineering Program, University of Colorado, Boulder, CO 80309, USA*

[6]*Renewable and Sustainable Energy Institute, National Renewable Energy Laboratory and University of Colorado, Boulder, CO 80309, USA*

[*]*Email: ivan.smalyukh@colorado.edu*

# Supplementary Note

**A concept of elastic dipole in the framework of nematostatics**

We assume that the far-field LC director $\mathbf{n}_0$ is oriented parallel to the $z$ axis. Then, at large distances from a particle that causes elastic distortions, the director field may be approximated as $\mathbf{n}(\mathbf{r}) \approx \left(n_1, n_2, 1 - O(n_1^2, n_2^2)\right)$, and the Frank-Oseen bulk free energy in the one elastic constant approximation is given by

$$F_{FO} \approx \int_V d^3 r \left( \sum_{i=1,2} (\nabla n_i)^2 + O(n_i^4) \right). \qquad \text{(Supp. Eq. 1)}$$

Consequently, equilibrium transverse director distortions $n_i$ obey the Laplace equation

$$\Delta n_i = 0. \qquad \text{(Supp. Eq. 2)}$$

In analogy to the electrostatics, the solutions for $n_i$ can be written in term of multipoles

$$n_i = q_i \frac{1}{r} + \sum_{\alpha=1}^{3} d_{i\alpha} \frac{r_\alpha}{r^3} + \ldots . \qquad \text{(Supp. Eq. 3)}$$

Rotational covariance of the director field around the $z$ axis requires that $q_i \equiv 0$. $\mathbf{d}_i \equiv (d_{ix}, d_{iy}, d_{iz})^T$ is the $i^{th}$ dipole moment determined by the following surface integral

$$d_{i\alpha} = -\frac{1}{4\pi} \int_S d^2 s\, n_i N_\alpha , \qquad \text{(Supp. Eq. 4)}$$

where $S$ is some arbitrary particle-enclosing surface (with the topology of a sphere), and $N_\alpha$ is the $\alpha^{th}$ component of the unit inward vector normal to $S$. $\mathbf{d}_i$ do not transform as vectors and can be referred to as components of dipole dyad.[1] The dyadic form of the elastic dipole is related to the specific reference frame (with $\mathbf{n}_0 \parallel Oz$) used here. In a general reference frame the elastic dipole is characterized by a three-dimensional second rank tensor.

In Ref. [1], it was shown that the 2×3 matrix, whose first and second rows are formed by the components of $\mathbf{d}_1$ and $\mathbf{d}_2$ respectively, may be presented as a direct sum $\mathbf{D} \oplus \mathbf{d}_3$ of the two-dimensional second rank tensor $\mathbf{D}$ and the two-dimensional vector $\mathbf{d}_3$. Furthermore, there exists a special reference frame $O_0$ in which $\mathbf{D}$ takes the following form

$$\mathbf{D}_0 = d \begin{pmatrix} 1 & 0 \\ 0 & 1 \end{pmatrix} + \Delta \begin{pmatrix} 1 & 0 \\ 0 & -1 \end{pmatrix} + C \begin{pmatrix} 0 & 1 \\ -1 & 0 \end{pmatrix}. \qquad \text{(Supp. Eq. 5)}$$

If we assume that in reference frame $O_0$ $\mathbf{d}_3 = (\gamma, \gamma')^T$, then the dipolar dyad has the following form

$$\mathbf{d}_{01} = (d + \Delta, C, \gamma)^T , \qquad \text{(Supp. Eq. 6)}$$

$$\mathbf{d}_{02} = (-C, d - \Delta, \gamma')^T . \qquad \text{(Supp. Eq. 7)}$$

Therefore, the general elastic dipole is described by three parameters $d$, $\Delta$, $C$ and a 2D vector $(\gamma, \gamma')^T$.

According to Ref.[1], the parameters $d$, $\Delta$, and $C$ are called isotropic, anisotropic and chiral dipole strengths, respectively. Vector $\mathbf{d}_3$ is called the longitudinal dipole, which corresponds to the elastic dipole dyad components along $\mathbf{n}_0$. The discussion in the main text of this article adopts this terminology when considering elasticity-mediated colloidal interactions between chiral microparticles with the same or opposite handedness.

**Supplementary References**

1. Pergamenshchik, V. M. & Uzunova, V. A. Dipolar colloids in nematostatics: Tensorial structure, symmetry, different types, and their interaction. *Phys Rev. E* **83**, 021701 (2011).

## Supplementary Figures

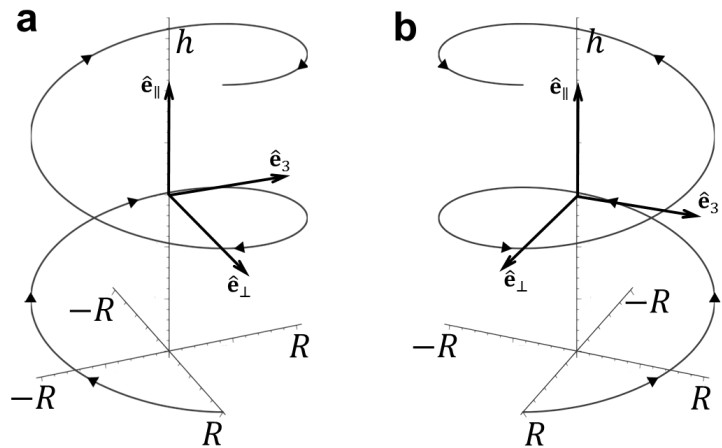

**Supplementary Figure 1 | Geometry of colloidal springs and helices. a**, Left-handed and **b** right-handed helices/springs defined by equation (2). Arrows indicate the direction of the increase of the parameter $\tau \in [0, 4\pi]$. In each case right-handed orthonormal reference frame ($\hat{\mathbf{e}}_\parallel$, $\hat{\mathbf{e}}_\perp$, $\hat{\mathbf{e}}_3$) is attached to the chiral particle. The frame is useful for describing particle orientation relative to each other and the far-field director. The spatial span of the particles along the $\hat{\mathbf{e}}_\parallel$ axis is denoted $h$ while that along the $\hat{\mathbf{e}}_\perp$, $\hat{\mathbf{e}}_3$ axes is $2R$.

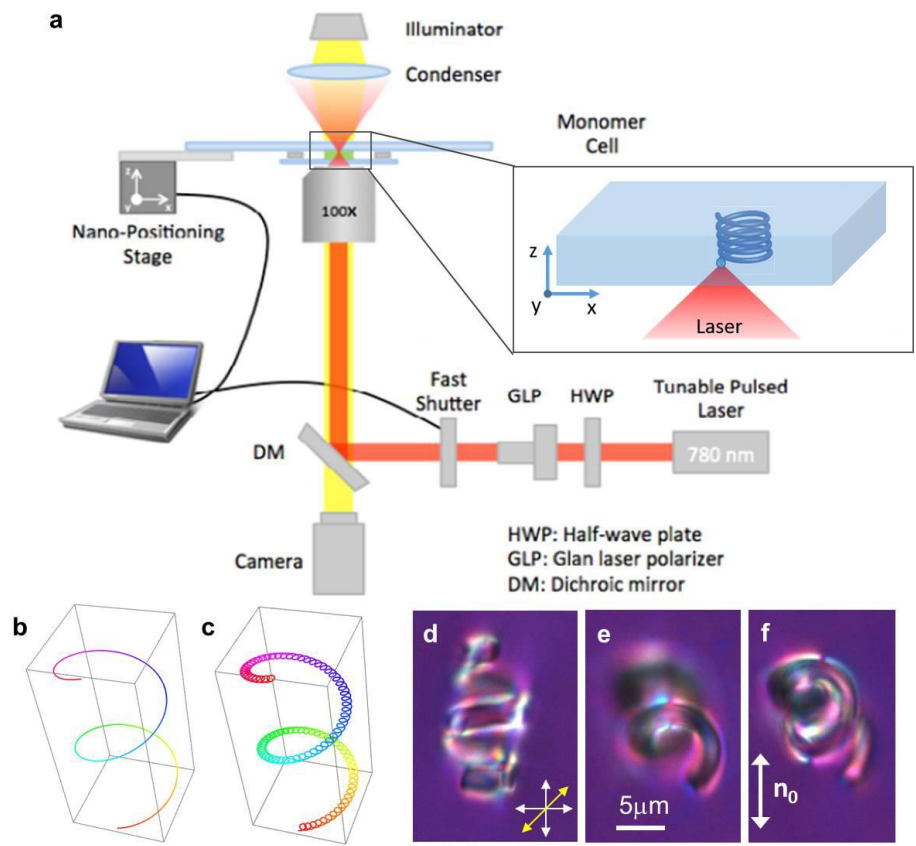

**Supplementary Figure 2 | Two-photon photopolymerization of chiral microparticles. a**, A highly simplified schematic of the home-built two-photon photopolymerization setup showing its key components. A labview-based computer software controls timing between the fast shutter and the nano-positioning stage in order to "draw" desired structures of the photopolymerized solid microparticles. The inset shows a schematic representation of the photopolymerization cell with monomer and photoinitiator in the form of a droplet sandwiched between a glass slide and a microscope coverslip. Pre-programmed translation of the focal point of the focused laser beam within the cell yields chiral microparticles with different handedness. **b**, **c**, Examples of laser beam translation trajectories used to photopolymerize **b**, thin and **c**, thick chiral microparticles. The looped spiraling trajectory shown in **c**, allows for making the spring-shaped particles mechanically more robust and with a well defined circular cross-section of the spring's tube. **d**, **e**, **f**, Examples of photopolymerized chiral microsprings viewed from different 3D perspectives in an aligned nematic sample (with the far-field director $\mathbf{n}_0$ shown in **f**, using the thick white double arrow). The crossed polarizers and the slow axis of the 530 nm phase retardation plate corresponding to the polarizing micrographs are shown in **d**, using thin white and yellow double arrows, respectively.

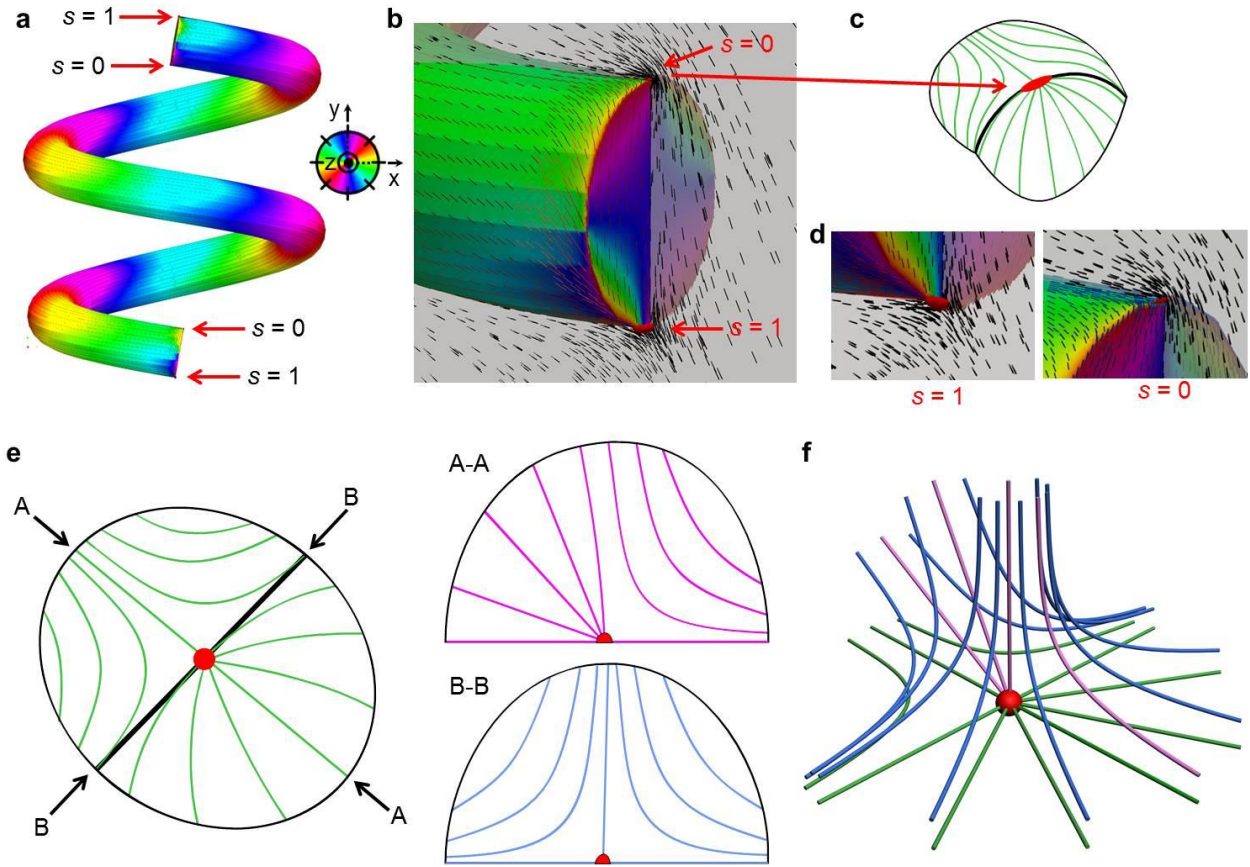

**Supplementary Figure 3 | Surface defects with a neutral ($s=0$) topological hedgehog charge. a**, A left-handed spring with color-coded azimuthal orientations of $\mathbf{n}(\mathbf{r})$ on the particle surface with respect to $\mathbf{n}_0$ aligned along the $z$-axis according to the color scheme shown in the inset. **b**, Detailed $\mathbf{n}(\mathbf{r})$ around the bottom end of the particle shown in **a**. **c,** Schematic representation of the top edge of the bottom end shown in **b** and $\mathbf{n}_s(\mathbf{r})$ (green lines) around a defect's singular core (a red ellipsoid) located on the particle's edge. **d**, Detailed $\mathbf{n}(\mathbf{r})$-configurations around the defects with $s=1$ and $s=0$. **e**, Reconstruction of the director field configuration: the left-side illustration is a schematic of the flattened edge and $\mathbf{n}_s(\mathbf{r})$ on the surface of the particle shown in (**c**); the right-side schematics are $\mathbf{n}(\mathbf{r})$ (magenta and blue lines) around a defect's singular core in the bulk around the defect in the vertical planes A-A and B-B marked in the left image by black arrows. **f**, 3D schematic showing director field $\mathbf{n}(\mathbf{r})$ (green, magenta and blue lines) around the surface defect in (**c, d, e**).

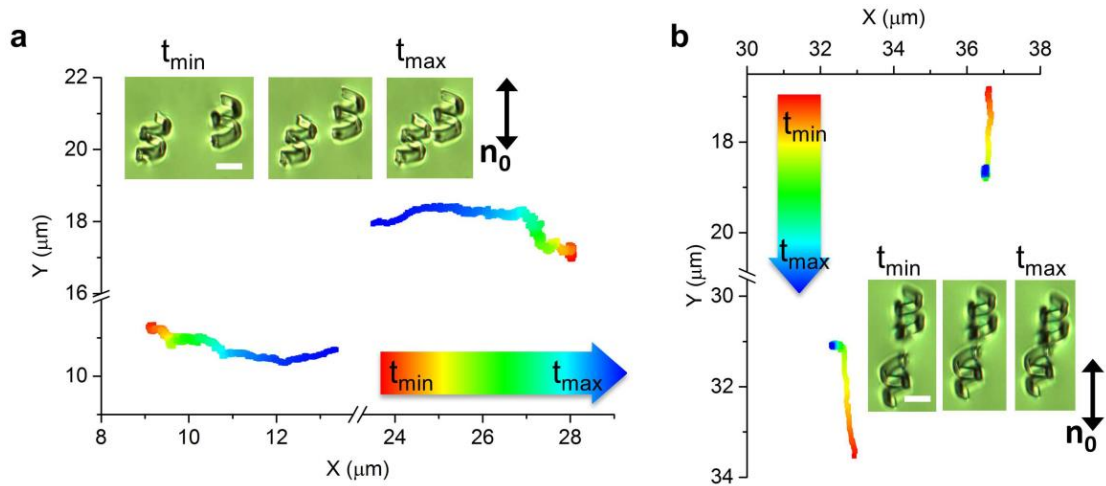

**Supplementary Figure 4 | Colloidal pair interactions of springs released from the laser tweezers with the center-to-center separation vectors initially oriented at an angle with respect to $n_0$. a**, **b**, Time--color-coded trajectories illustrating the colloidal interaction of **a**, like-handed and **b**, oppositely-handed colloidal microsprings. The bright field micrographs in the insets show relative positions of particles at different elapsed times. The colors scales of elapsed time (insets) depict the color-coded time counted from the moment of releasing particles from the laser traps ($t_{min}$) till the moment when they approach each other $t_{max}$ ($t_{max}$-$t_{min}$=95s in **a** and $t_{max}$-$t_{min}$=5s in **b**), as shown in the bright field micrographs in the insets; the far-field director $n_0$ is along the vertical edges of micrographs, as shown using thick double arrows. Scale bars are 5 μm.

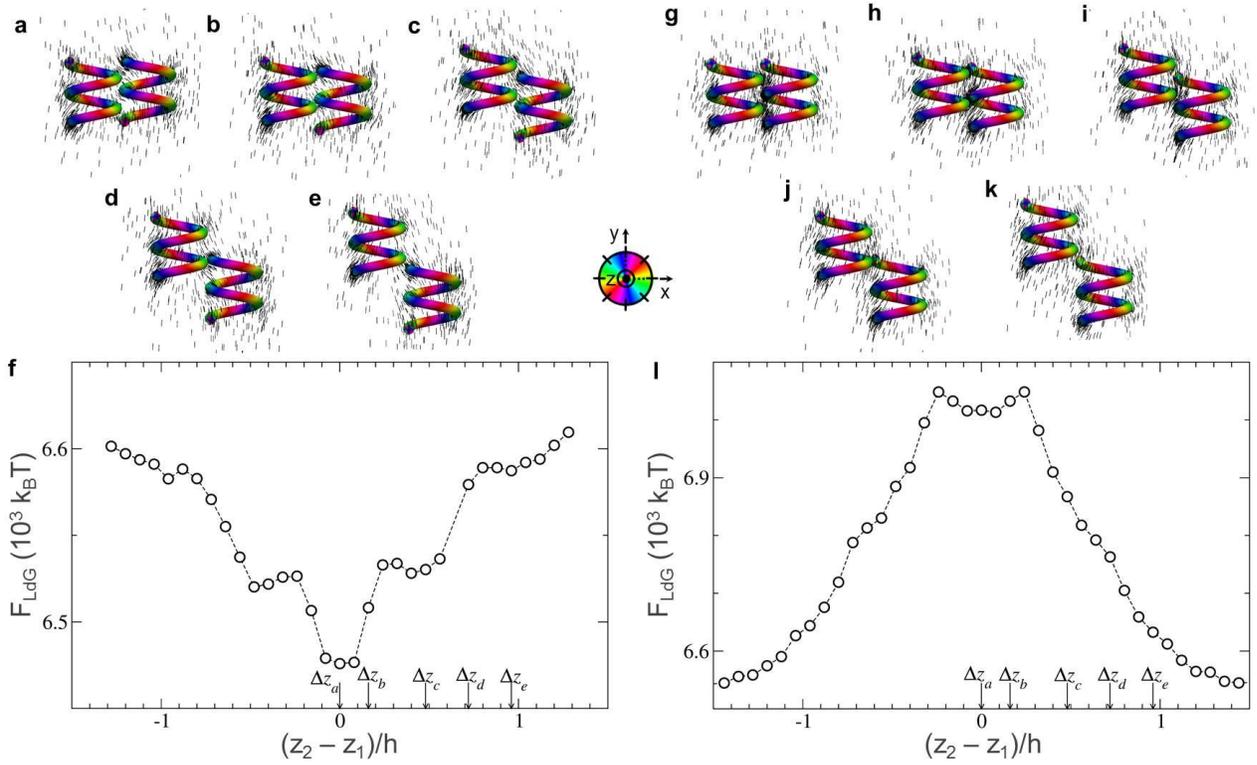

**Supplementary Figure 5 | Free energy of interaction between (a-f) oppositely-handed and (g-l) like-handed microsprings oriented along $n_0$ and with their centers "elastically" locked on lines along $n_0$. a-e** and **g-k** show **n(r)** around the springs using colors on the particle surfaces and black rods in the bulk LC. The colors on the particle surfaces depict azimuthal orientations of **n(r)** with respect to $n_0$ according to the color scheme in the middle. **f, l** Numerically calculated Landau-de Gennes free energy as a function of the component of $\vec{d}$ along $n_0$, $d_{||} = z_2 - z_1$, for (**f**) oppositely-handed and (**l**) like-handed springs. The perpendicular component $d_\perp$ of $\vec{d}$ is fixed at 1.2$R$. The helix axes $\hat{e}_{||}$ are parallel to $n_0$.

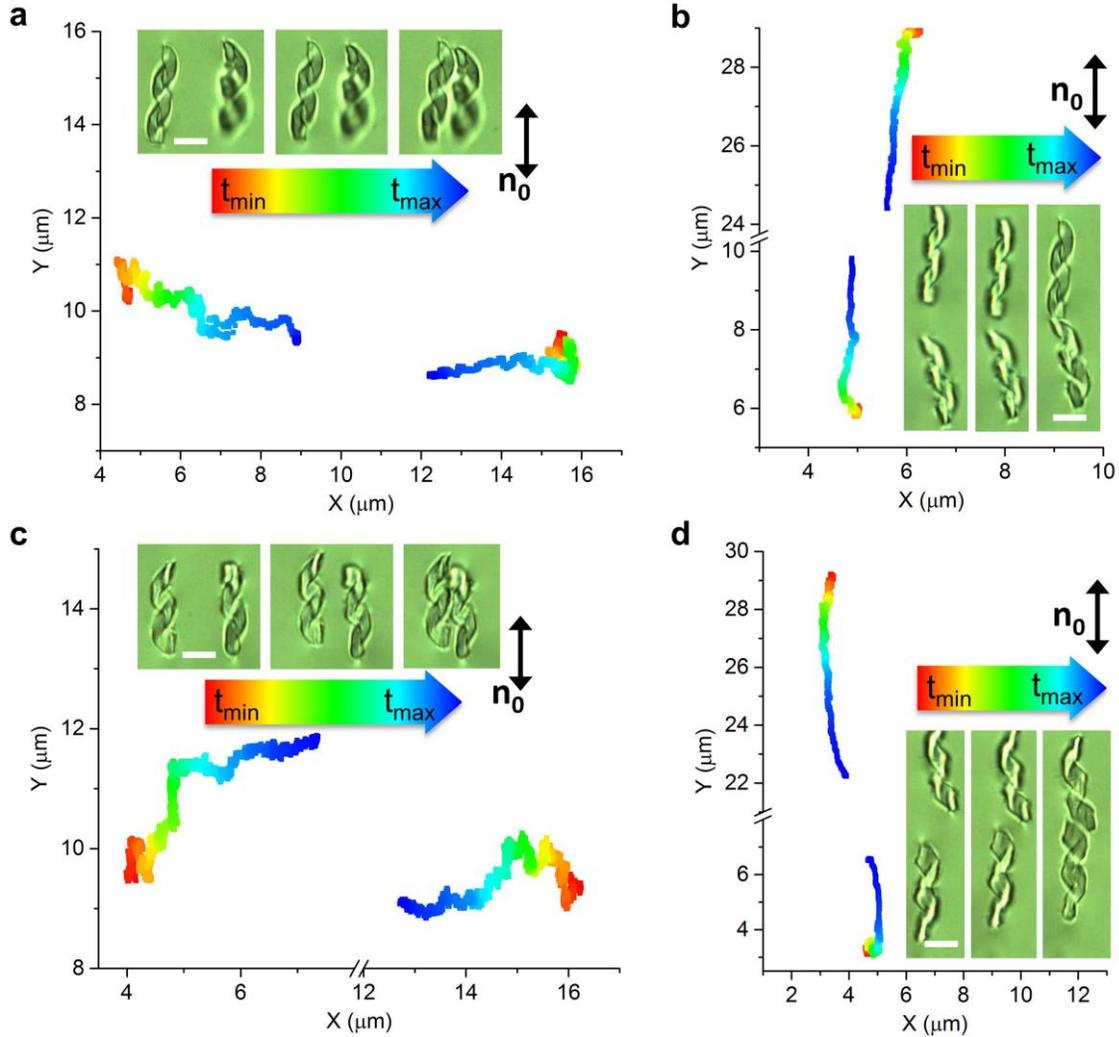

**Supplementary Figure 6 | Chirality-dependent elastic pair interactions between colloidal helices. a**, **b**, Time-coded trajectories of attraction of oppositely-handed helices initially separated so that the center-to-center separation vector $\vec{d}$ is **a**, orthogonal and **b**, parallel to $\mathbf{n}_0$. **c**, **d**, Time-coded trajectories of attraction of like-handed helices initially separated so that $\vec{d}$ is **c**, orthogonal and **d**, parallel to $\mathbf{n}_0$. The insets show color scales of elapsed time counted from the moment of releasing particles from laser traps ($t_{min}$) till the moment when they approach each other $t_{max}$ ($t_{max}$-$t_{min}$=170s in **a**, $t_{max}$-$t_{min}$=35s in **b**, $t_{max}$-$t_{min}$=150s in **c**, and $t_{max}$-$t_{min}$=70s in **d**). The bright field micrographs in the insets were obtained at (left) $t_{min}$, (right) $t_{max}$ and (middle) at an intermediate time. The direction of $\mathbf{n}_0$ in the micrographs is shown using black double arrows. Scale bars are 5 μm.

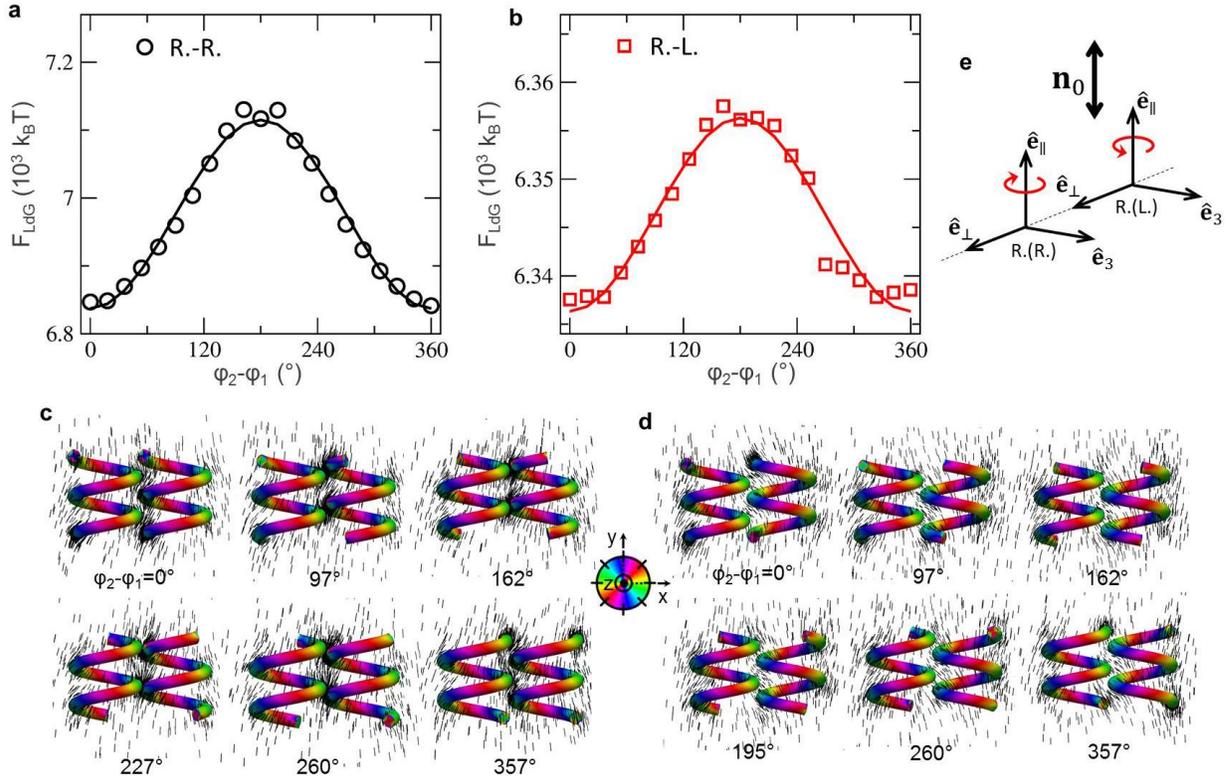

**Supplementary Figure 7 | Free energy of chirality-dictated rotation of microsprings about their $\hat{\mathbf{e}}_\parallel$ axes, which are aligned along $\mathbf{n}_0$.** Free energy cost of elastic distortions induced by the rotation of the **a,** like-handed and **b,** oppositely-handed particles about their $\hat{\mathbf{e}}_\parallel$ axes in the opposite directions by equal amounts; $\varphi_2-\varphi_1$ is the angle between the orientations of particle's $\hat{\mathbf{e}}_\perp$ vectors, as shown schematically in **e**. The scenter-to-center distance is $d=1.2R$ and the angle between the center-to-center vector $\vec{d}$ and $\mathbf{n}_0$ is fixed at 90°. The solid lines are fitting curves based on equation $A+B\cos(\varphi_2-\varphi_1)$, where $A$, $B$ are fitting parameters. $\mathbf{n}(\mathbf{r})$ around the like-handed **c,** and oppositely-handed **b,** springs is shown using colors on the particle surfaces and black rods in the bulk LC.

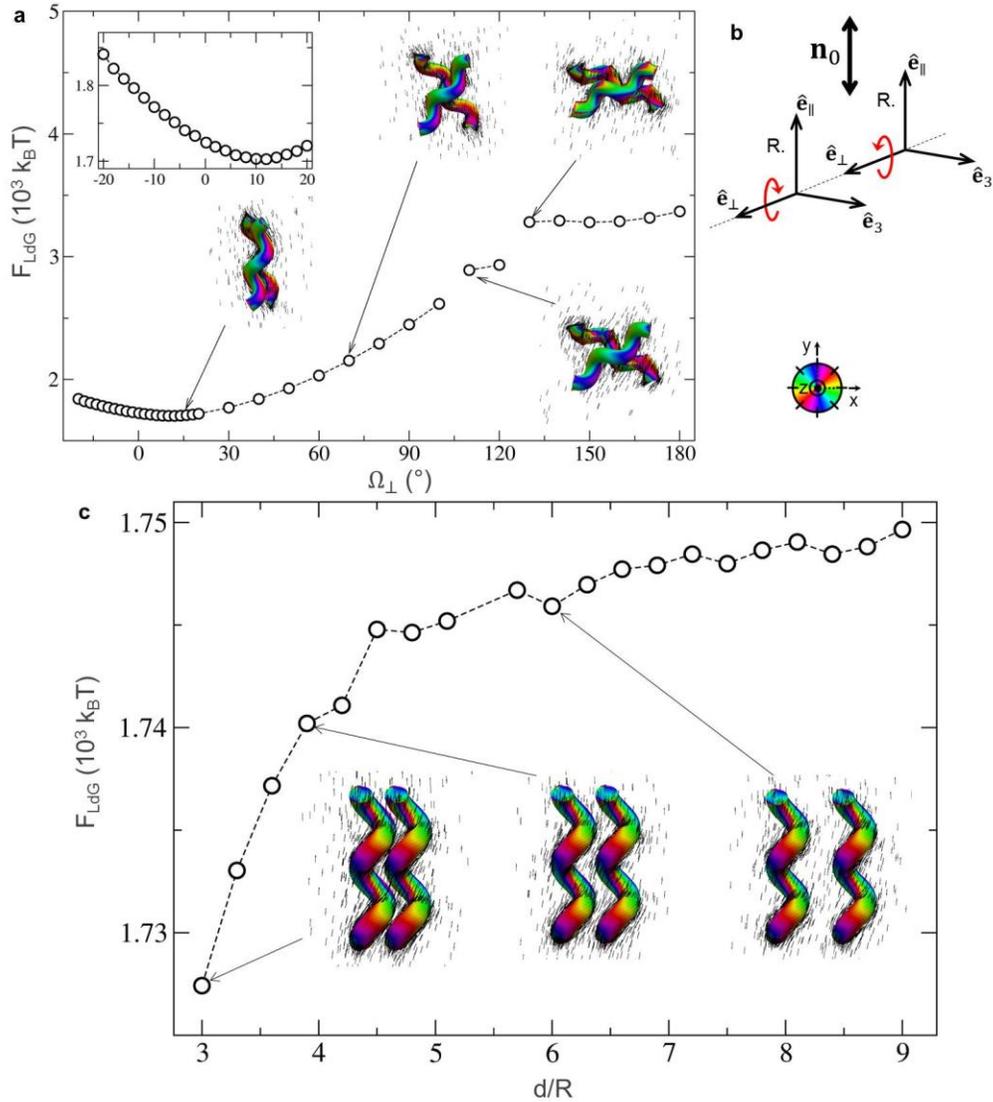

**Supplementary Figure 8 | Free energy associated with pair interaction of like-handed helices.**
**a**, Numerically calculated Landau-de Gennes free energy as a function of the angle $\Omega_\perp$ between the axes $\hat{\mathbf{e}}_\parallel$ of two like-handed helices; center-to-center distance $d=0.29R$, $\Psi=90°$. The helix axes rotate about the particles' $\hat{\mathbf{e}}_\perp$ vectors in opposite directions by equal amounts, as schematically shows in **b**. **c**, Numerically calculated Landau-de Gennes free energy as a function of the relative center-to-center distance $d$ between two like-handed single helices at $\Psi=90°$; the helix axes are parallel to $\mathbf{n}_0$. The insets in **a** and **c** show $\mathbf{n}(\mathbf{r})$ around the helices depicted using colors on particle surfaces and black rods in the LC bulk.

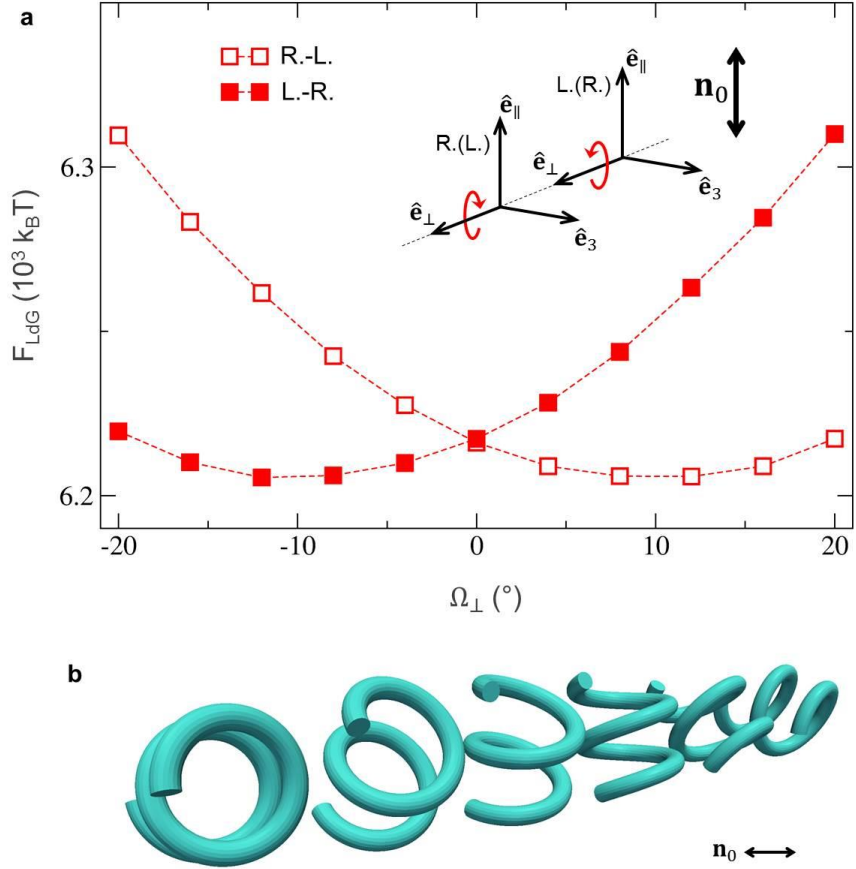

**Supplementary Figure 9 | Dependence of the free energy on the twisting direction for oppositely handed springs. a,** Numerically calculated Landau-de Gennes free energy as a function of the angle $\Omega_\perp$ between the spring axes $\hat{\mathbf{e}}_\parallel$ of opposite-handed springs: $d=1.625R$, $\Psi=90°$. The spring axes $\hat{\mathbf{e}}_\parallel$ rotate about the particles' $\hat{\mathbf{e}}_\perp$ axes in opposite directions by equal amounts, as schematically shown in the inset. **b**, An example of self-assembly of chiral colloidal spring particles, where colloidal springs in a nematic LC under confinement that forces alignment of $\vec{d}$ along $\mathbf{n}_0$ dispersed at large number densities self-assemble into helical structures by continuously rotating around an axis parallel to $\mathbf{n}_0$. Such confinement could be realized in cylindrical or rectangular capillaries with tangential anchoring for director and cross-section dimensions only slightly larger than the size of colloidal springs.